\documentclass[a4paper,10pt]{sigwebnewsletter}

\usepackage{graphicx}
\usepackage{amsmath}
\usepackage{enumitem}
\usepackage{svg}
\title{Nipping in the Bud: Detection, Diffusion and Mitigation of Hate Speech on Social Media}
\author{Tanmoy Chakraborty AND Sarah Masud \\Dept. of CSE, IIIT Delhi, \{tanmoy,sarahm\}@iiitd.ac.in}

\newsletterQuarter{Autumn}
\newsletterYear{2007}
\begin{abstract}
Since the proliferation of social media usage, hate speech has become a major crisis. Hateful content can spread quickly and create an environment of distress and hostility. Further, what can be considered hateful is contextual and varies with time. While online hate speech reduces the ability of already marginalised groups to participate in discussion freely, offline hate speech leads to hate crimes and violence against individuals and communities. The multifaceted nature of hate speech and its real-world impact have already piqued the interest of the data mining and machine learning communities. Despite our best efforts, hate speech remains an evasive issue for researchers and practitioners alike. This article presents methodological challenges that hinder building automated hate mitigation systems. These challenges inspired our work in the broader area of combating hateful content on the web. We discuss a series of our proposed solutions to limit the spread of hate speech on social media.
\end{abstract}

\begin{document}
\maketitle
\section{Introduction}
Digital platforms are now becoming the de-facto mode of communication. Owing to diverse cultural, political, and social norms being followed by the users worldwide, it is extremely challenging to set up a universally accepted cyber norms\footnote{\url{https://bit.ly/3mwbzQq}}. Compounding this complexity with the issue of online anonymity \cite{Suler2004}, the cases of predatory and malicious behaviour have increased with Internet penetration. Users may (un)intentionally spread harm to other users via spam, fake reviews, offensive, abusive posts, hate speech and so on. This article mainly focuses on hate speech on social media platforms. 

\begin{figure}
    \centering
    \includegraphics[width=0.85\textwidth]{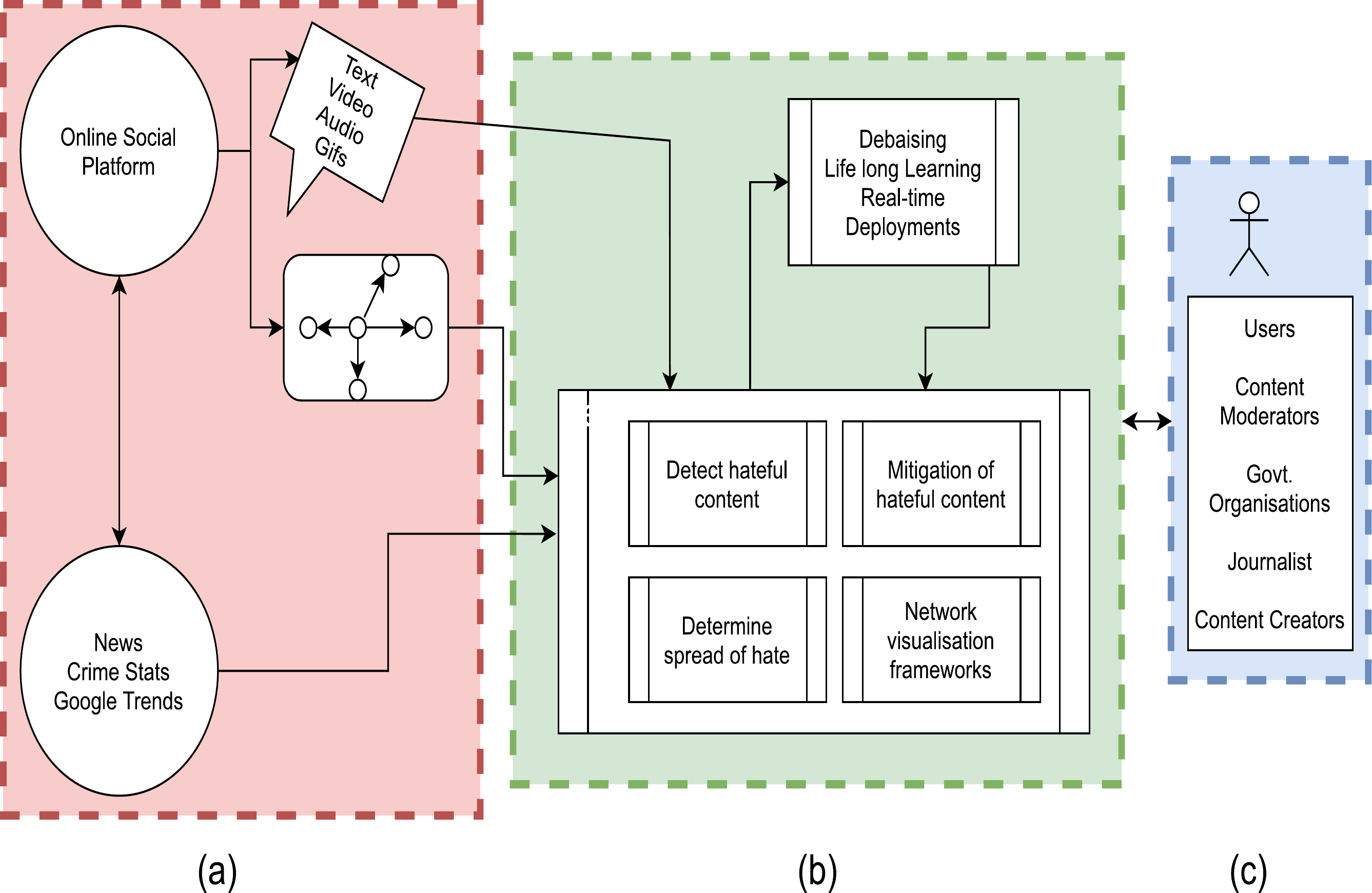}
    \caption{A framework for analysing and mitigating hate speech consists of the following: (a) Input signals made of `endogenous signals' (obtained from textual, multi-modal and topological attributes within the platform) and `exogenous signals' obtained from real-world events. (b) The input signals are curated to develop models for detecting hateful content and its spread; once detected, we can have a method to counter/mitigate hateful content. Since online platforms also involve large-scale interactions of users and topics, we require a framework for visualising the same. Once trained, these models can be deployed with feedback incorporating debasing and life-long learning frameworks. (c) Users, content moderators, organisations, and other stakeholders are at the receiving end of the framework. Their feedback and interactions directly impact the deployed systems.}
    \label{fig:overall_hate_framework}
\end{figure}

United Nations Strategy and Plan of Action defined hate speech as ``{\em any kind of communication in speech, writing or behaviour, that
attacks or uses pejorative or discriminatory language with reference to a person or a group on the basis of who they are, in other words, based on their religion, ethnicity, nationality,
race, colour, descent, gender or other identity factor }.''\footnote{\url{https://bit.ly/32psGwv}} 
%\textit{Hate speech can be defined as content that targets a person or group based on their attributes like gender, sexual orientation, race, entity, etc}}.
Across interactions, what can be considered hateful varies with geography, time, and social norms. However, the underlying intent to cause harm and bring down an already vulnerable group/person by attacking a personal attribute can be considered a standard point for defining hate speech. It leads to a lack of trust in digital systems and reduces the democratic nature of the Internet for everyone to interact freely \cite{Stevens2021-zs}. Further exposure to hateful content impacts the mental health of the people it is targeted at, and those who come across such content \cite{Saha2019}. Thus, we need to develop a system to help detect and mitigate hate speech on online social platforms. This paper discusses existing challenges, shortcomings and research directions in analysing hate speech. While not exhaustive, we hope our corpus of work will help researchers and practitioners develop systems to detect and mitigate hate speech and reduce its harmfulness on the web. A generic framework for analysing and mitigating hate speech can be understood via Figure \ref{fig:overall_hate_framework}.

\section{Existing Challenges}
The area of hate speech poses multiple challenges \cite{MacAvaney2019} for researchers, practitioners and lawmakers alike. 
These challenges make it difficult to implement policies at scale. This section briefly discusses some of these issues that inspire our body of work.
\label{sub:challenges}
\begin{enumerate}
\item\label{item:C1} \textbf{C1: What is considered as hateful?} Different social media platforms have different guidelines to manage what communication is deemed acceptable on the platform. Due to the lack of a sacrosanct definition of hate speech, researchers and practitioners often use hate speech as an umbrella term to capture anti-social behaviours like toxicity, offence, abuse, provocation, etc. It makes determining the malicious intent of an online user challenging. A  blanket ban on users for a single post is, therefore, not a viable long-term solution \cite{Johnson2019}.
\item\label{item:C2} \textbf{C2: Context and subjectivity of language:}
Human language is constantly evolving, reflecting the zeitgeist of that era. Most existing hate speech detection models fail to capture this evolution depending on a manually curated hate lexicon. On the other hand, offenders constantly find new ways to evade detection \cite{10.1145/3270101.3270103}. Apart from the outdated hate-lexicons lies the problem of context (information about the individual's propensity for hate and the current worldview). Determining isolated incidents of hate speech is difficult even for a human with world knowledge. 
\item\label{item:C3} \textbf{C3: Multifaceted nature of communication on the Internet:} Online communication exists in varying forms of text, emoji, images, videos or a combination of them. These different modalities provide varying cues about the message. In this article, we talk specifically about memes as a source of harmful content \cite{NEURIPS2020_1b84c4ce,cOriol}. Memes are inherently complex multi-modal graphics, accommodating multiple subtle references in a single image. It is difficult for machines to capture these real-life contexts holistically. 
\item\label{item:C4} \textbf{C4: Lack of standardised large-scale datasets for hate speech:} A random crawling of online content from social media platforms is always skewed towards non-hate \cite{hateoffensive,ICWSM1817909}. Due to the content-sharing policies of social media platforms, researchers cannot directly share text and usually release a set of post ids for reproducibility. By the time these ids are crawled again, the platform moderators have taken down explicit hate speech, and the respective user accounts may be suspended. Additionally, due to changes in the context of hate speech, text once annotated as hate may need to be rechecked and reannotated, leading to a lack of standardised ground-truth for hate speech \cite{Kovcs2021}.
\item\label{item:C5} \textbf{C5: Multilingual nature of hate speech:} All the challenges discussed above compound for the non-English system. Natural language processing models consume text as data and determine usage patterns of words and phrases. It is hard to develop statistically sound systems for low-resource and code-mixed content without training on large-scale data \cite{10.1145/3457610}. Take, for example, the case of collecting and annotating datasets under code-mixed settings. It is hard to train systems to detect which word is being spoken in which language. Consider, for example, a word spelt out as ``main", which means ``primary" in English, and in ``I, myself" in Hindi. Depending on the language of the current word, the meaning conveyed by a code-mixed sentence can change. 
\end{enumerate}

\section{Research Questions and Proposed Solutions}
\label{sec:prop}
\subsection{RQ1 --  Following the challenges C\ref{item:C2} and C\ref{item:C4}, can we predict the spread of hate on Twitter?}
\label{sec:rq1}
\textbf{Background}. For the experiments discussed in this section, we use textual and topological content extracted from the Twitter platform. A tweet is the smallest unit of communication in Twitter. A user can include text, images, URLs, hashtags in the tweet. Once posted, other users who follow the said users can observe the new tweet on their timeline. Upon exposure, a user can retweet, with or without appending anything to it or comment to start a thread. Each tweet is associated with a unique tweet id, each user with its unique user id. Using a combination of a tweet and user ids, we can use the Twitter APIs to crawl tweets and followers of a user holding a public account.

\subsubsection{\bf RQ1.1 -- Predicting retweeting activity of hateful users} Conversations on any online platform reflect events happening in the real-world (exogenous signals) and vice-versa. Capturing these signals can help us better understand which users are likely to participate in which conversation. In our work \cite{9458789}, we explore the theme of topic-dependent models for analysing and predicting the spread of hate. We crawled a large-scale dataset of tweets, retweets, user activity history, and follower networks, comprising more than $41$ million unique users. We also crawled $300k$ news articles. It was observed that hateful tweets spread quickly during the initial hours of their posting, i.e., users who engage in malicious content intend to propagate their hurtful messages as far and wide as quickly possible. These observations are in line with hateful behaviour on other platforms like Gab \cite{10.1145/3292522.3326034}.
Additionally, it was observed that users' propensity to engage in seeming hateful hashtags varies across different socio-political topics. Based on these observations, we proposed a model which, given a tweet and its hate markers (ratio of hateful tweets and hateful retweets on the user's timeline), along with a set of topical and exogenous signals (news title in this case), predicts which followers of the said user are likely to retweet hateful posts. The motive behind using exogenous signals is to incorporate the influence of external events on a user's posting behaviour \cite{10.1145/3394486.3403251}. Interestingly, we observed that existing information diffusion models that do not consider capturing any historical context of a user or incorporate exogenous signals perform comparably on non-hateful cascades but fail to capture diffusion patterns of hateful users. It happens because the dataset is skewed towards non-hateful cascades. In the absence of latent signals, only topological features are not enough to determine the spread of hate.
    
\subsubsection{\bf RQ1.2 -- Predicting hatefulness of Twitter reply threads} In another set of work \cite{10.1145/3447548.3467150}, we define the problem of forecasting hateful replies – the aim is to anticipate the hate intensity of incoming replies, given the source tweet and a few of its initial replies. The hate intensity score constitutes a prediction probability of a hate speech classifier and the mapping of words from hate-lexicon\footnote{\url{https://hatebase.org/}}, which had a score manually curated for each hate word. Over a dataset of $1.5k$ Twitter threads, we observed that the hatefulness of the source tweet does not correlate with the hatefulness of the replies eventually received. Hate detection models on individual tweets could not predict the inflexion going from benign to hateful. By modelling the thread as a series of discrete-time hate intensities over a moving window, we proposed a ``blind state deep model'' %\cite{NEURIPS2018_5cf68969}
that predicts the hate intensity for the next window of the reply thread. Here blind means one does not need to specify the underlying function, and the deep state captures the non-linearity. Our experimental results found that the proposed model is more robust than baselines when controlled for the underlying hate speech classifier model, the length of the reply thread and the type of source tweet considered (fake, controversial, regular, etc.). This robustness is expected from a model deployed for environments as dynamic and volatile as the social media platforms.

\subsection{RQ2 -- Following challenges C\ref{item:C3}, can harmful memes be a precursor for conveying hate?}
\label{sec:rq2}
\textbf{Background.} With the proliferation of memes, they are now being used to convey harmful sentiments. Owing to their subtle messaging, they easily bypass automatic content flagging. Offensive memes that target individuals or organisations based on personal attributes like race, colour, and gender are deemed hateful. On the other hand, harmful memes are a rather border category. These memes can be offensive \cite{suryawanshi-etal-2020-multimodal}, hateful, abusive or even bullying\footnote{\url{https://wng.org/sift/memes-innocent-fun-or-internet-bullying-1617408938}} in nature. Additionally, harm can be intended in multiple ways like -- loss of credibility of the target entity or disturbing mental peace and self-confidence of the target entities. In the next set of works, we propose some benchmark datasets and models to detect the harmfulness of online memes as well as their targeted entities.

\subsubsection{\bf RQ2.1 -- Harmful meme benchmark dataset}
\label{sec:rq2.1}
To narrow down the scope of harmful memes, we begin by selecting the topic of COVID-19. The variety of content covered by this topic and its social relevance in the current times make it an apt choice for our work \cite{pramanick-etal-2021-detecting}. From Google Image Search %\footnote{\url{https://www.google.com/imghp}} 
and public pages of Instagram %\footnote{\url{https://www.instagram.com/}} 
and Reddit,%\footnote{\url{https://www.reddit.com/}}, 
we curated a dataset of $3.5k$ memes. This dataset is named as \textit{HarMeme}. In the first step of the annotation, we labelled the memes as `very harmful', `partially harmful' or `harmless'. In the second step, we additionally annotated the harmed target entity as either an `individual' (e.g., Barack Obama), `organisation' (e.g., WHO), `community' (e.g., Asian-American), or `society' at large. On this dataset, we evaluated various baselines under uni-modal and multi-modal settings. Even our best performing method, a multi-modal architecture with an accuracy of $81.36\%$, failed to reach the human benchmark of $90.68\%$. This benchmark was annotated by a group of expert annotators (separate from those who participated in crowd annotation). For our second problem of detecting the target of harm, we again found that the best performing multi-modal framework falls short of the human benchmark ($75.5\%$ vs $86.01\%$ accuracy). These differences in accuracy highlight the non-trivial nature of the harmful meme detection task. 

\subsubsection{\bf RQ2.2 -- Detecting harmful memes under multi-modal setting}
In a subsequent work \cite{pramanick-etal-2021-momenta-multimodal}, we extended the above \textit{HarMeme} dataset to include US Politics as a topic as well. Following the same annotation process, we ended up with two harmful meme datasets, called {\em Harm-C} and {\em Harm-P}, covering the Covid-19 and US-politics, respectively. We then proposed a multi-modal framework that encodes image and text features along with image attributes (background, foreground, image attributes) obtained from the Google Vision API. These features are fused via inter and cross-modal attention mechanisms and trained under a multi-task setting. Compared to the best performing multi-modal baseline, our proposed model improved $\approx1.5\%$ accuracy in both tasks. However, the gap between the human benchmark (as described in Section \ref{sec:rq2.1}) and the proposed method is still significant. It begs the question of adding more signals to capture context. Additionally, we performed ablation of domain transfer where we trained on one set of harm memes and tested on others. The proposed model incorporating pretrained encoding from CLIP \cite{radford2021learning} showed improved transferability compared to baselines.

\subsection{RQ3 -- Following challenges C\ref{item:C1} and C\ref{item:C4}, can we use cues from anti-social behaviour to predict hatefulness?}
\label{sec:rq3}
Our recent work \cite{Sengupta2021} explored the detection of various offensive traits under a code-mixed Hinglish (Hindi+English) setting. We combined our dataset from existing Hinglish datasets on aggression \cite{kumar-etal-2018-aggression}, hate \cite{bohra-etal-2018-dataset}, humour \cite{khandelwal-etal-2018-humor}, sarcasm \cite{swami2018corpus} and stance \cite{swami2018englishhindi}. Since a single data source consisting of all the above categories does not exist, we developed pseudo-labels for each task and at each data point. Our ablation studies showed that using pseudo-labels under a multi-task setting improved the performance across all predictive classes. We further looked at microscopic (word-level) as macroscopic (task/category-level) causality that can help explain the model's prediction. To perform word-level dependency on a label, we generated a \textit{causal importance score} \cite{NEURIPS2020_92650b2e} for each word in a sentence. It captures the change in the confidence level of prediction in a sentence's presence vs absence of that word. We observed that the mean of the importance score lies around zero for all categories, with low variance for the overtly aggressive and hate classes.
On the other hand, we observed a higher variance in importance score for sarcasm, humour, and covert aggression. It follows from puns and wordplay that contextually impacts the polarity of words. Further, we employed the Chi-square test between all pairs of offensive traits to determine how the knowledge of a prior trait impacts the prediction for the trait under consideration. We observed that an overtly aggressive text has $25\%$ higher chances of being hateful than other classes, and knowing that it is not aggressive lowers its chance of being hateful by $50\%$. Therefore, prior knowledge about the aggressiveness of a text can impact posterior probabilities of the text being hateful.

\begin{table}[!t]
\centering
\scalebox{0.8}{
\begin{tabular}{|p{0.25\textwidth}|p{0.325\textwidth}|p{0.325\textwidth}|}
\hline
\textbf{Research Question} &
  \textbf{Proposed Solution} &
  \textbf{Dataset Curated} \\ \hline
RQ1.1  Can we predict hateful rewteets? &
  Exogenous attention modeling. &
  Tweets, retweets, user history, \& follower networks ($41M$ unique users). $300k$ news articles. Manually annotated $17k$ tweets for hate/non-hate. \\ \hline
RQ1.2 Can we predict hatefulness of reply threads? &
  Blind state deep model. &
  Tweet reply threads consisting of $1.5k$ threads with average length $200$ per thread. \\ \hline
RQ2.1 Can we curate a meme dataset for type and target of harm? &
  Benchmark existing uni and multi-modal frameworks for harmful memes. &
  $3.5k$ memes annotated for harmful or not, as well as target of harm. Human benchmarks against the annotated dataset are also provided. \\ \hline
RQ2.2 Can we bring the performance of harmful meme detection models closer to human benchmarks? &
  Inter and Intra-modal attention in a multi-task multi-modal framework. &
  $\approx 7.5k$ memes annotated for harmful or not, as well as target of harm. Human benchmarks against the annotated dataset are also provided. \\ \hline
RQ3 Can offensive traits lead to hate? &
  Pseudo-labelled multi-task framework to predict the offensive traits. &
  Combined from five existing datasets of offensive trait predictions in Hinglish. \\ \hline
\end{tabular}}
\caption{Summary of research questions, methods and curated datasets discussed in this article.}
\label{tab:rq_summary}
\end{table}

\section{Future Work}
We summarize the entire discussion in Table \ref{tab:rq_summary}.
Off-the-shelf hate speech classifiers were employed in diffusion and intensity prediction models. However, existing hate speech classifiers have been reported to be biased against the very communities they hope to help \cite{sap-etal-2019-risk}. Therefore, incorporating debased models and proposing such techniques for code-mixed settings can be a direction for future work. Further, other forms of unintended bias like political bias have been studied very scantly and require additional investigations \cite{wich-etal-2020-impact}. Apart from the problem of bias is the issue of static hate-lexicons. We need robust %\cite{rottger-etal-2021-hatecheck} 
and explainable %\cite{Mathew_Saha_Yimam_Biemann_Goyal_Mukherjee_2021} 
systems that evolve. % \cite{qian-etal-2021-lifelong}.
Many regional languages on social media go unnoticed until some socio-political unrest surges in the region, e.g., Facebook's inability to timely moderator content in Mayanmar\footnote{\url{https://www.reuters.com/investigates/special-report/myanmar-facebook-hate/}}. Like the multilingual transformer models, research in hate speech calls for developing transfer learning systems \cite{10.1145/3457610} that can contextually capture target entities and hateful phrases across domains. %\cite{nozza-2021-exposing}. 
Other multi-media contents like gifs and short clips are also worth exploring for analysing harmful content. The gap in human benchmarks and our best performing multi-modal frameworks shows that detecting harmful memes requires additional context beyond visual and textual features \cite{pramanick-etal-2021-momenta-multimodal}. Knowledge graphs \cite{10.1145/3430984.3431072}, and diffusion patterns are potential signals to incorporate in future studies. As both knowledge graphs and diffusion cascades are hard to analyse and comprehend, various tools have been proposed in visualising these systems at scale \cite{sahnan2021diva,ilievski2020kgtk}. 

Meanwhile, studies have shown that the best counter to hate speech is not banning content but producing more content that sensitises the users about hate speech \cite{Hangartner2021}. In this regard, reactive and proactive counter-speech strategies need to be worked out \cite{chaudhary2021countering}. While we have primarily spoken about tackling hate speech from a computer science perspective, a topic as historically rich and sensitive as hate speech requires multi-disciplinary efforts. Theories from sociology and physiology might help researchers and practitioners better understand the emergence and spread of hate from a socio-cultural perspective. Additionally, by involving stakeholders like representatives from minority communities, journalists, content moderators, we will be able to deploy solutions that are human-centric and not data-centric.

% \section{ACKNOWLEDGEMENTS}
% We would like to thank Shivam Sharma and Shivani Kumar for their valuable inputs.

\bibliographystyle{sigwebnewsletter}
\bibliography{instructions}

\begin{biography}
\noindent\textbf{Sarah Masud}
is a Prime Minister PhD Scholar at the Laboratory for Computational Social Systems (LCS2),  IIIT-Delhi, India. Within the broad area of social computing, her work mainly revolves around hate speech detection and diffusion. She regularly publishes papers in top conferences including SIGKDD and ICDE. Before joining her PhD, she worked on Recommendation Systems at Red Hat. \\
Homepage: \url{https://sara-02.github.io}.

\noindent\textbf{Tanmoy Chakraborty} is an Assistant Professor of Computer Science and a Ramanujan Fellow at IIIT-Delhi. Prior to this, he was a postdoctoral researcher at University of Maryland, College Park. He completed his PhD as a Google PhD scholar at IIT Kharagpur, India. 
%His PhD thesis got the best thesis awards from Indian National Academy of Engineering (INAE), IBM Research and Xerox Research. 
His research group, LCS2, broadly works in the areas of social network analysis and natural language processing, with special focus on cyber-informatics and adversarial data science. He has received several prestigious awards including Faculty Awards from Google, IBM, Accenture. He was also a DAAD visiting faculty at Max Planck Institute for Informatics. 
%He has been working with industries including Google, IBM, Wipro, Accenture, Logically, Flipkart, Hike. He is also involved in mentoring several tech startups. 
He has recently authored a textbook on ``Social Network Analysis''. 
%He has been in organizing committee of NAACL'21, PAKDD'21, and in PC/SPC/AC committee of AAAI, ACL, EMNLP, SIGKDD, The WebConf, IJCAI, etc. He has served as a (guest) editor of four special issues in journals.   \\
Homepage: \url{http://faculty.iiitd.ac.in/~tanmoy}; Lab page: \url{http://lcs2.iiitd.edu.in/}.

\end{biography}
\end{document}